\documentclass[a4paper]{article}

\usepackage{INTERSPEECH2019}
\usepackage{amsmath, multirow, array, verbatim}
\usepackage[inline]{enumitem}

\title{Many-to-Many Voice Conversion with Out-of-Dataset Speaker Support}
\name{Gokce Keskin, Tyler Lee, Cory Stephenson, Oguz H. Elibol}
\address{Intel AI Lab, Santa Clara, CA, USA}
\email{gokce.keskin@intel.com}

\begin{document}

\maketitle
\begin{abstract}
We present a Cycle-GAN based many-to-many voice conversion method that can convert between speakers that are not in the training set. This property is enabled through speaker embeddings generated by a neural network that is jointly trained with the Cycle-GAN. In contrast to prior work in this domain, our method enables conversion between an out-of-dataset speaker and a target speaker in either direction and does not require re-training. Out-of-dataset speaker conversion quality is evaluated using an independently trained speaker identification model, and shows good style conversion characteristics for previously unheard speakers. Subjective tests on human listeners show style conversion quality for in-dataset speakers is comparable to the state-of-the-art baseline model. 
\end{abstract}

\section{Introduction}
\label{sec:intro}
Converting the voice of a source speaker to a target style has been studied in the context of voice conversion and speaker de-identification \cite{Stylianou:1998,Jin:2009}. Gaussian Mixture Models (GMM) are among the first successful voice conversion techniques \cite{Stylianou:1998, Toda:2007}. These models are trained on \textit{parallel} datasets, which contain the same utterances (e.g., words or sentences) from source and target speakers. Parameter adaptation techniques can extend parallel-data trained GMMs to speakers with \textit{non-parallel} data \cite{Mouchtaris:2006}. A Long Short-Term Memory (LSTM) based many-to-one voice conversion method trained on non-parallel data has been proposed, but this method relies on an automatic speech recognition (ASR) system that is trained on labeled data \cite{Sun:2016}. 

More recently, deep-learning based methods such as Variational Auto-Encoders (VAE) and Cycle-Consistent Generative Adversarial Networks (Cycle-GAN) have been used to perform voice conversion using solely non-parallel data \cite{Kaneko:2017, Kameoka:2018, Hosseini:2018, Fang:2018, Kameoka:2018_2, Hsu:2017}. Typically, source speaker features are extracted using a vocoder such as WORLD \cite{Masanori:2016} or STRAIGHT \cite{Kawahara:1999}, selected features are converted to the target speaker's features, and finally the target voice is built in audio domain using the converted features. Alternatively, linear spectrograms can be used as features for voice conversion \cite{Hosseini:2018}, since the converted spectrogram can be mapped back to the waveform domain using the Griffin-Lim algorithm \cite{Griffin:1984}.

\begin{figure}[t!]
\begin{center}
\includegraphics[width=0.6\columnwidth]{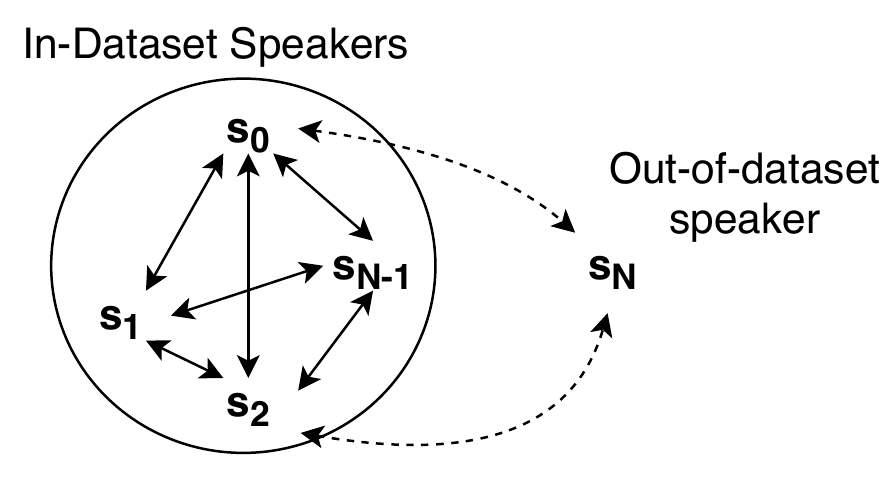}
\end{center}
\caption{In addition to in-dataset, many-to-many  voice conversion, proposed methodology can convert from/to out-of-dataset speakers that it was not trained on.}
\label{fig:voice_conv}
\end{figure}

In this paper, we propose a many-to-many voice conversion framework that uses speaker embeddings and a Cycle-GAN. Many-to-many conversion is defined as converting the voice of any desired speaker in a source set of speakers to the style of a target speaker (Fig. \ref{fig:voice_conv}). A convolutional neural network (CNN) based \textit{generator} is used for the conversion (Sec. \ref{ssec:proposed}). A CNN-based \textit{feature extractor}, jointly trained with the Cycle-GAN, learns speaker-specific embeddings (features). Embeddings are subsequently used to condition the output of the generator to the style of the desired speaker. 
One unique attribute that sets the proposed methodology apart from prior work is the ability to perform conversion between an out-of-dataset speaker $\pmb{s_{N}}$ and any in-dataset speaker $\pmb{s_i} (0\leq i<N)$ in either direction \textit{without re-training} (Fig. \ref{fig:voice_conv}). This property is enabled by the learned speaker embeddings. Our results demonstrate that feature extractors trained on a diverse set of source speakers can generalize well and generate embeddings for speakers that were not in the training set, enabling voice conversion for out-of-dataset speakers (Sec\ref{ssec:ood_comp}). In subjective tests on humans, in-dataset style conversion quality for the proposed model is comparably rated to the state-of-the-art baseline model(Sec. \ref{ssec:baseline_comp}).

\section{METHODOLOGY}
\label{sec:methodology}
\subsection{Proposed Methodology}
\label{ssec:proposed}
Cycle-GANs were originally proposed as a domain conversion technique in images where collecting parallel data between the desired domains is either expensive or impossible, such as converting from photographs to paintings \cite{Zhu:2017}. As an appealing parallel-data-free method, it was adapted for one-to-one and many-to-many voice conversion where collecting time-aligned parallel data between speakers is costly \cite{Kaneko:2017, Kameoka:2018}. The proposed methodology in this paper builds on a prior Cycle-GAN-based method in \cite{Kameoka:2018} by adding a feature extractor block to enable out-of-dataset speaker conversion and combining the two separate discriminators into a single block.

\begin{figure}[t!]%
\centering
\includegraphics[width=0.8\columnwidth]{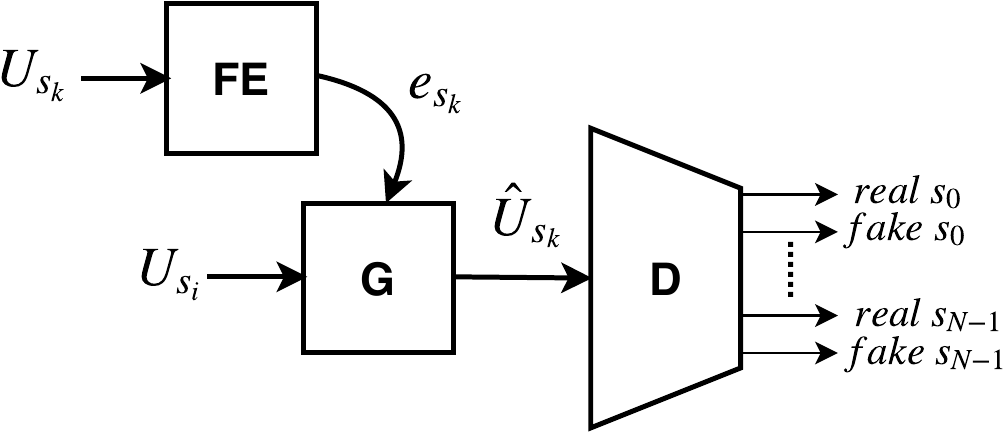}%
\caption{Proposed model consists of generator $\pmb{G}$, feature extractor $\pmb{FE}$ and discriminator $\pmb{D}$.
$\pmb{G}$ converts input utterance $U_{s_{j}}$ from speaker $\pmb{s_{j}}$ to the style of speaker $\pmb{s_{k}}$ using embedding $e_{s_{k}}$. Embedding is produced by $\pmb{FE}$ from sample utterance $U_{s_{k}}$ of speaker $\pmb{s_{k}}$. 
Converted utterance $\hat{U}_{s_{k}}$ has the content of $U_{s_{j}}$, but the style of $\pmb{s_{k}}$. 
$\pmb{D}$ is trained to classify genuine utterance $U_{s_{k}}$ as $real \; s_k (r^\prime_{s_k})$, and fake utterance $\hat{U}_{s_{k}}$ as $fake \; s_k (f^\prime_{s_k})$. $\pmb{FE}$ and $\pmb{G}$ are trained to generate $\hat{U}_{s_{k}}$ such that $\pmb{D}$ will classify it as $r^\prime_{s_k}$.
When trained on a diverse dataset of speakers, $\pmb{FE}$ can provide embeddings for out-of-dataset speakers, enabling voice conversion from/to an out-of-dataset speaker without re-training any of the components.
}%
\label{fig:cycle_gan1}%
\end{figure}

Fig. \ref{fig:cycle_gan1} shows the basic components of the proposed many-to-many voice converter. $U_{s_{j}}$ is the mel-spectrogram of a genuine utterance from source speaker $\pmb{s_{j}}$, and $U_{s_{k}}$ is the mel-spectrogram of a genuine utterance from the target speaker $\pmb{s_{k}}$. $U_{s_{j}}$ and $U_{s_{k}}$ are unaligned and do not contain the same sentence, and potentially do not even share a single word. Feature extractor $\pmb{FE}$ produces speaker embedding $e_{s_{k}}$, latent representation of the style of speaker $\pmb{s_k}$. Generator $\pmb{G}$ converts $U_{s_{j}}$ to $\hat{U}_{s_{k}}$ using $e_{s_{k}}$, trying to mimic the style of the target speaker while keeping the content in $U_{s_{j}}$. Discriminator $\pmb{D}$ takes in an utterance and classifies it into one of $2N$ outputs, where $N$ is the number of speakers in the training set. $\hat{U}_{s_{k}}$ is converted to the audio domain using Griffin-Lim algorithm \cite{Griffin:1984}.

$\pmb{D}$ is trained to classify genuine utterance $U_{s_{j}}$ as $real \; s_{j}$. In other words, $r^\prime_{s_j}$ output is maximized, and all other outputs of $\pmb{D}$ are minimized (Fig. \ref{fig:cycle_gan1}). Similarly, when presented with the generated (fake) utterance $\hat{U}_{s_{k}}$,  output $f^\prime_{s_{k}}$ is maximized.  This is achieved by adjusting the parameters of $\pmb{D}$ to minimize the loss $\mathcal{L}_{D}$:
\begin{equation} 
\label{eq:Dloss}
\mathcal{L}_{D} = -\Big( 
                      \mathbb{E}[r_{s_{i}}\log(r^\prime_{s_i})\big]
                    + \mathbb{E}[f_{s_{i}}\log(f^\prime_{s_i})\big] 
				\Big)
\end{equation}
where $0\leq i < N$. $r_{s_i}$ is a binary indicator variable and is set to one only when the input of $\pmb{D}$ is a real utterance from speaker $\pmb{s_i}$. $f_{s_i}$ is an indicator variable that denotes the input is a generated utterance in the style of speaker $\pmb{s_i}$. During training, $\pmb{D}$ is presented with real and fake utterances for all speakers in the training set.

There is a single $\pmb{FE}$ and a single $\pmb{G}$; together they perform all conversions. They are trained to generate $\hat{U}_{s_{k}}=\pmb{G}(U_{s_{j}}, e_{s_{k}})$ such that $\pmb{D}$ will be fooled to label the generated utterance as $real \; s_{k}$, i.e. output $r^\prime_{s_k}$ is maximized. In practice, this is achieved by tuning the parameters of $\pmb{FE}$ and $\pmb{G}$ to minimize $\mathcal{L}_{G}$:
\begin{equation} 
\label{eq:Gloss}
\mathcal{L}_{G} = - 
\mathbb{E}[f_{s_{i}}\log(r^\prime_{s_{i}})]
\end{equation}

$\pmb{FE}$ and $\pmb{G}$ are trained for all in-dataset speakers jointly and are shared for all speakers. Initially, fake utterances generated by $\pmb{G}$ with the help of $\pmb{FE}$ are low quality and $\pmb{D}$ can easily discriminate the fakes from genuine utterances. As training progresses, $\pmb{G}$ learns to mimic the target speaker better, and $\pmb{D}$ learns to find better ways to distinguish fakes from real utterances. 

Although $\pmb{G}$ and $\pmb{FE}$ might eventually learn to generate utterances that sound very much like the target speaker, $\mathcal{L}_{D}$ and $\mathcal{L}_{G}$ do not guarantee that the content in $\hat{U}_{s_{k}}$ and $U_{s_{j}}$ will match. For example, $\pmb{G}$ might generate an utterance that says ``horse" and sounds very much like the target speaker, but keeps generating ``horse" even if $\pmb{s_j}$ contains ``cat" or ``chicken".

In Cycle-GAN based methods, content preservation is enforced by \textit{cycle consistency loss}. The input utterance $U_{s_{j}}$ is first converted to $\hat{U}_{s_{k}}=\pmb{G}(U_{s_{j}}, e_{s_{k}})$, an utterance in the style of $\pmb{s_{k}}$. This utterance is passed through the generator a second time to generate utterance $U^\prime_{s_{j}}=\pmb{G}(\hat{U}_{s_{k}}, e_{s_{j}})$, which has the style of $\pmb{s_{j}}$. The output of the cycle, $U^\prime_{s_{j}}$, should ideally be the same as $U_{s_{j}}$. This is enforced through the L1-loss $\mathcal{L}_{cycle}$:
\begin{equation} \label{eq:cycle_loss}
\mathcal{L}_{cycle} = \mathbb{E}\big[\lVert U_{s_{j}} - U^\prime_{s_{j}}  \rVert_{1}\big]
\end{equation}

$\pmb{G}$ and $\pmb{FE}$ are trained to minimize $\mathcal{L}_{cycle}$, with source and target speakers randomly chosen during training. 

\subsection{Comparison to Prior Work}
\label{ssec:related}
The ability to convert between a source speaker and an out-of-dataset speaker without re-training is a key difference between the proposed methodology and all prior work in this domain \cite{Kaneko:2017, Kameoka:2018, Hosseini:2018, Fang:2018, Kameoka:2018_2, Hsu:2017}. Converters described in \cite{Kaneko:2017, Fang:2018, Hsu:2017} can only perform one-to-one conversion; and \cite{Hosseini:2018} only performs domain conversion between female and male speakers, rather than conversion between specific speakers.

A Cycle-GAN based voice conversion method presented in \cite{Kameoka:2018} is the most closely related prior work to our model. The proposed feature extractor block used to generate speaker embeddings, as opposed to the \textit{training attribute} one-hot vector in \cite{Kameoka:2018}, is the key differentiator that enables out-of-dataset speaker support. Additionally, we demonstrate voice conversion between more than 290 speakers (Sec. \ref{ssec:ood_comp}), a significantly more complex task than the four speakers presented in \cite{Kameoka:2018}. Our model has a single discriminator as opposed to two separate discriminators, enabling a slightly simpler architecture. We use mel-spectrograms for conversion instead of features generated by a separate vocoder (Sec. \ref{ssec:preprocess}).

\section{IMPLEMENTATION DETAILS}
\label{sec:implementation}

\subsection{Data Preprocessing}
\label{ssec:preprocess}
Our model uses mel-spectrograms computed with Librosa and parameters \texttt{$n\_fft$=512, $hop\_length$=32, $n\_mels$=128, $fmin$=40 and $fmax$=7900} \cite{librosa}. Per-frequency scaling is performed for each speaker by: 
\begin {enumerate*}[label=\itshape\alph*\upshape)]
\item Take a random subset of the audio files for the speaker and clip silence from selected files, \item Compute the log-magnitude mel-spectrograms, \item Compute the histogram for each frequency bin, \item Take $99.9^{th}$ percentile value for each frequency bin as the maximum allowed value ($max$) for that bin, and choose ($max-4$) as the minimum ($min$), \item Clip all values in the spectrogram to $[min, max]$ and scale to $[-1, +1]$.
\end {enumerate*} 
During training, pre-computed $min$ and $max$ values for each speaker are used to scale the log-magnitude mel-spectrograms.

\subsection{Dataset}
\label{ssec:dataset}
In Sec. \ref{ssec:baseline_comp}, we use four speakers from the Voice Conversion Challenge 2018 dataset to compare to the baseline for in-dataset speakers \cite{vcc2018}. In Sec. \ref{ssec:ood_comp}, training is performed on Librispeech train-clean-100 dataset for 251 speakers, approximately 25 minutes of total utterances per speaker \cite{Panayotov:2015}. Out-of-dataset speakers are chosen among 40 speakers in the dev-clean dataset. Transcriptions are not used in either of the two cases.

\subsection{Network Architecture}
\label{ssec:architecture}
\begin{figure}[t!]
\begin{center}
\includegraphics[width=0.9\columnwidth]{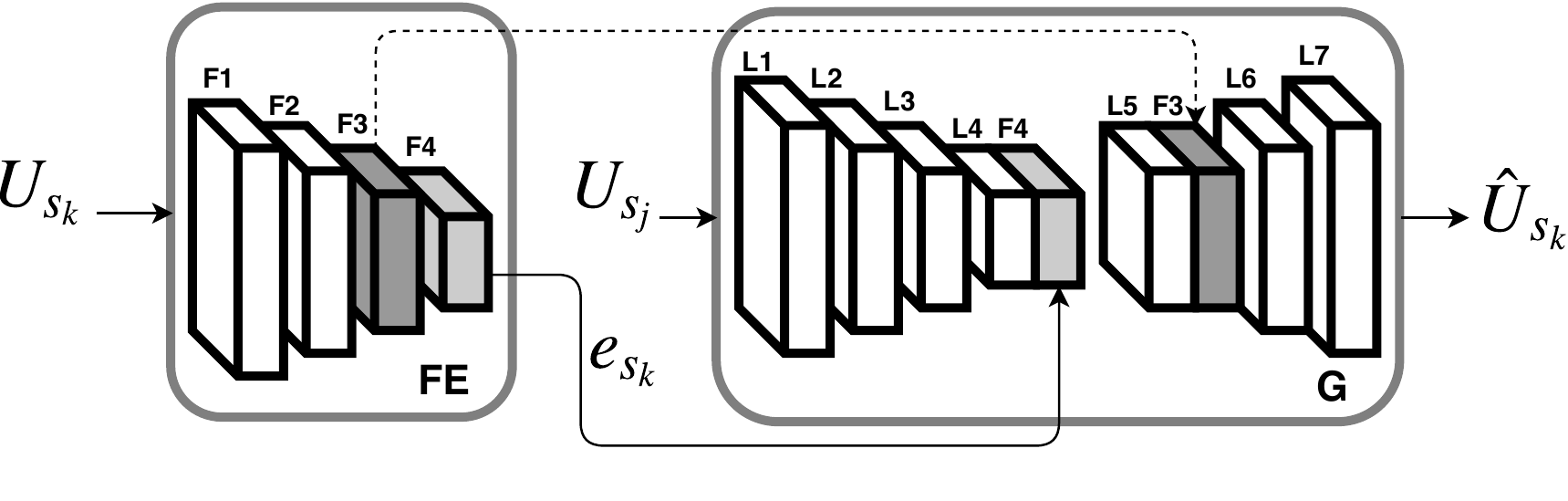}
\end{center}
\caption{Generator and feature extractor architectures. Each rectangular block (e.g., F1) is a two-layer CNN. Downsampling is performed by strided convolutions, upsampling is performed using sub-pixel and transposed convolutions. Outputs of F3 and F4 are mean pooled along time axis, then repeated to match the output dimensions of L5 and L4.}
\label{fig:architecture}
\end{figure}

Architecture of the many-to-many conversion path is given in Fig. \ref{fig:architecture}. Both $\pmb{G}$ and $\pmb{FE}$ are CNNs with 2D convolutions. Each unit (e.g., L1) consists of two convolutional layers, with desired downsampling performed at the second layer of the unit. $\pmb{FE}$ converts the mel-spectrogram of utterance $U_{s_k}$ from source speaker $\pmb{s_k}$ to the speaker-specific embedding $e_{s_k}$. An input receptive field of shape $1 \times 128 \times 64$ (channels, frequency bins, time steps) is mapped to an embedding of shape $1\times8\times8$ (CHW). Downsampling is performed by strided convolutions, and instance normalization layers are used to aid training \cite{Ulyanov:2016}. Layers F1, F2, L1 and L2 use gated convolutional units \cite{Dauphin2017}.

Downsampling path of $\pmb{G}$ shares the same CNN architecture as $\pmb{FE}$. $\pmb{G}$ combines the mel-spectrogram of utterance $U_{s_j}$ from speaker $\pmb{s_j}$ with embedding $e_{s_k}$ to generate the mel-spectrogram of $\hat{U}_{s_k}$, an utterance with the style of speaker $\pmb{s_k}$ and the content of $U_{s_j}$. Embedding $e_{s_k}$ is concatenated to the latent representation of $U_{s_j}$ in the bottleneck layer. F3 layer output from $\pmb{FE}$ is mean pooled along time axis, repeated to match the dimension of the first upsampling layer (L5) output, and then concatenated to the output of L5, resembling a U-Net architecture \cite{Ronneberger:2015}. Similarly, F4 output is mean pooled and repeated to match the output of L4. Upsampling is performed by transposed and sub-pixel convolutions \cite{Aitken:2017}.

Discriminator $\pmb{D}$ is implemented as a collection of three CNNs. All three look at 128 frequency bins of the input spectrogram with progressively wider time patches (32, 64 and 128 timesteps). Each has a linear layer at the end, followed by a softmax activation with $2N$ outputs as described in Sec. \ref{ssec:proposed}. Since silence does not contain speaker-identifying information, patches with signal power below a threshold are not sent to the discriminators.

\section{Results}
\label{sec:results}
\subsection{Comparison to Baseline}
\label{ssec:baseline_comp}
In-dataset conversion quality of the proposed methodology is compared to the most closely related work in \cite{Kameoka:2018}, state-of-the-art non-parallel converter based on Cycle-GANs. We were unable to locate the source code of this work and open-source versions from third parties \cite{songxiang, hujinsen} did not match the quality of the 12 conversion samples given by the original authors \cite{stargan_samples}. Published samples are from four unique (source, target) speaker pairs, with three audio clips from each pair. For a fair comparison, we use these published samples to evaluate the naturalness and style conversion quality of our method using subjective tests on human listeners. Four chosen pairs are (SM1$\rightarrow$SF1), (SM2$\rightarrow$SM1),  (SF1$\rightarrow$SF2), and (SF2$\rightarrow$SM2) from  \cite{vcc2018}. 

\subsubsection{Mean Opinion Score}
We evaluate the naturalness of the converted samples with the Mean Opinion Score (MOS) test. In each test, listeners are given a single audio file from one of three categories: 
\begin {enumerate*}[label=\itshape\alph*\upshape)]
\item A ground truth audio file from one of four speakers in the dataset (SF1, SF2, SM1, SM2)
\item An audio conversion output by the baseline
\item An audio conversion output by the proposed model.
\end{enumerate*}

Listeners are asked ``How natural is the speech in this audio clip?" and given five choices in a Likert scale: unnatural, somewhat unnatural, indifferent, somewhat natural, natural. Choices are mapped to integers from 1 (unnatural) to 5 (natural). The test is repeated by using audio files from all four speakers and conversions.

\begin{figure}[t!]
\begin{center}
\includegraphics[width=0.7\columnwidth]{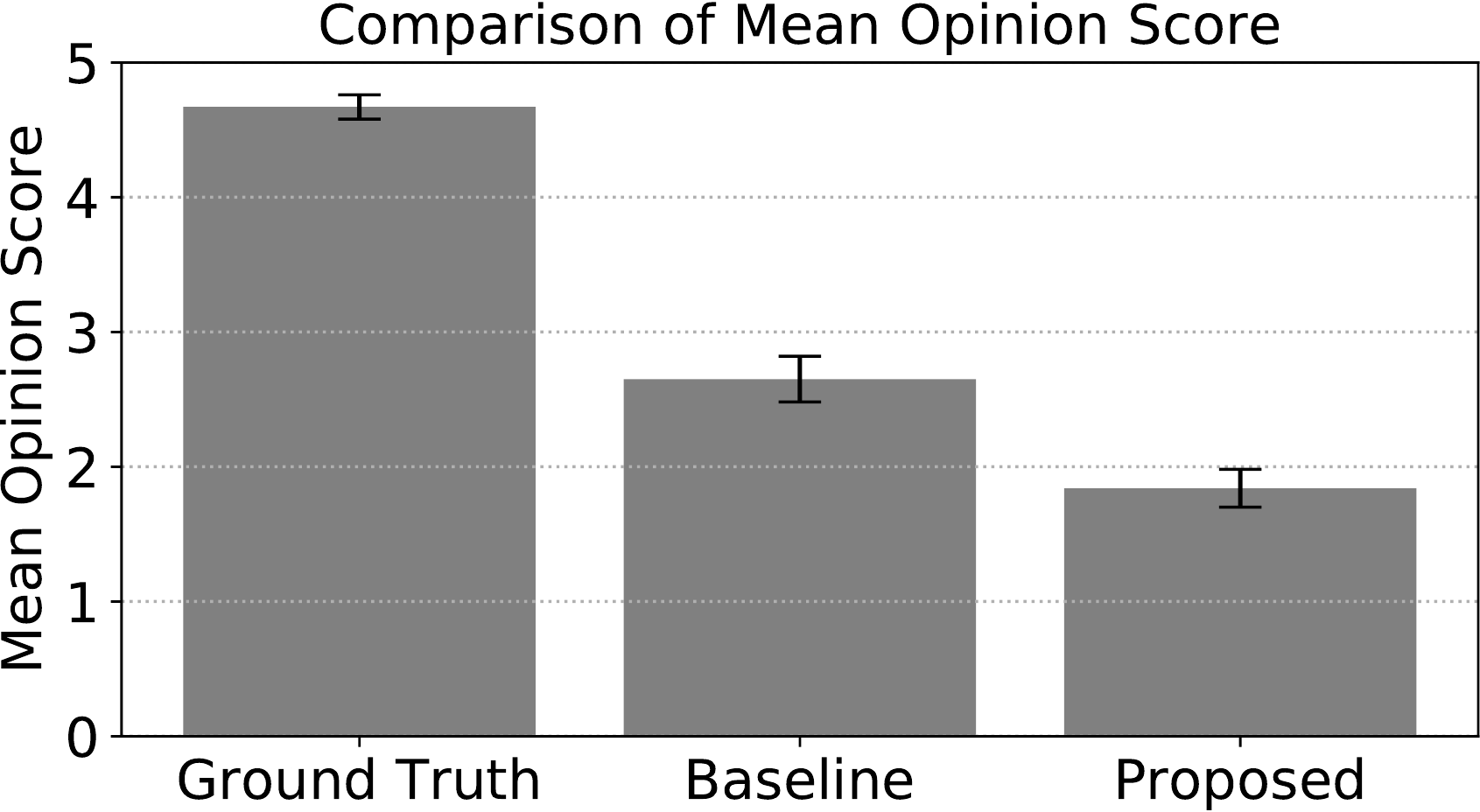}
\end{center}
\caption{Comparison of mean opinion scores (MOS) for utterances from ground truth (no conversion), baseline conversion and the proposed conversion. Higher scores mean listeners rated the given utterance more natural sounding. Ground truth results are reported as comparison to the two conversion models, along with 95\% confidence intervals. Proposed model is rated slightly lower, likely due to the Griffin-Lim inversion artefacts as opposed to the vocoder used in the baseline.}
\label{fig:results_mos}
\end{figure}

Fig. \ref{fig:results_mos} shows the comparison of MOS ratings from the baseline and proposed conversions to no conversion (ground truth audio files). Listeners rated the naturalness of the ground truth audio files almost at ``natural" level. Both conversion models are rated substantially below perfect conversion: Proposed model is rated slightly below "somewhat unnatural", whereas the baseline is slightly above. We surmise that the lower MOS scores of the proposed model is due to the artefacts introduced by Griffin-Lim algorithm when mel-spectrograms are rebuilt in raw audio domain. Baseline model uses a vocoder for audio reconstruction, which might be introducing fewer artefacts.

\subsubsection{Style Conversion Quality}
We evaluate the style conversion quality by comparing perceived similarity of conversion outputs to ground truth audio files from the intended target. Listeners are given two utterances: 
\begin {enumerate*}[label=\itshape\alph*\upshape)] 
\item A ground truth utterance from one of the four speakers in the dataset (e.g., SF1) 
\item Either another ground truth sample from this speaker (SF1), or an utterance converted to the style of this speaker by the baseline (output of SM1$\rightarrow$SF1 conversion), or the same conversion performed by the proposed model.
\end{enumerate*}

Listeners are asked ``How likely is that these two utterances are from the same speaker?" and given five choices in a Likert scale: unlikely, somewhat unlikely, neutral, somewhat likely, likely. Choices are mapped to integers from 1 (unlikely) to 5 (likely). The test is repeated with all four speaker pairs and from both conversion models. Comparing the perceived similarity of conversions to ground truth targets in a Likert scale gives more insight into the conversion models' capability when both models perform similarly good or bad in the conversion.

\begin{figure}[t!]
\begin{center}
\includegraphics[width=0.9\columnwidth]{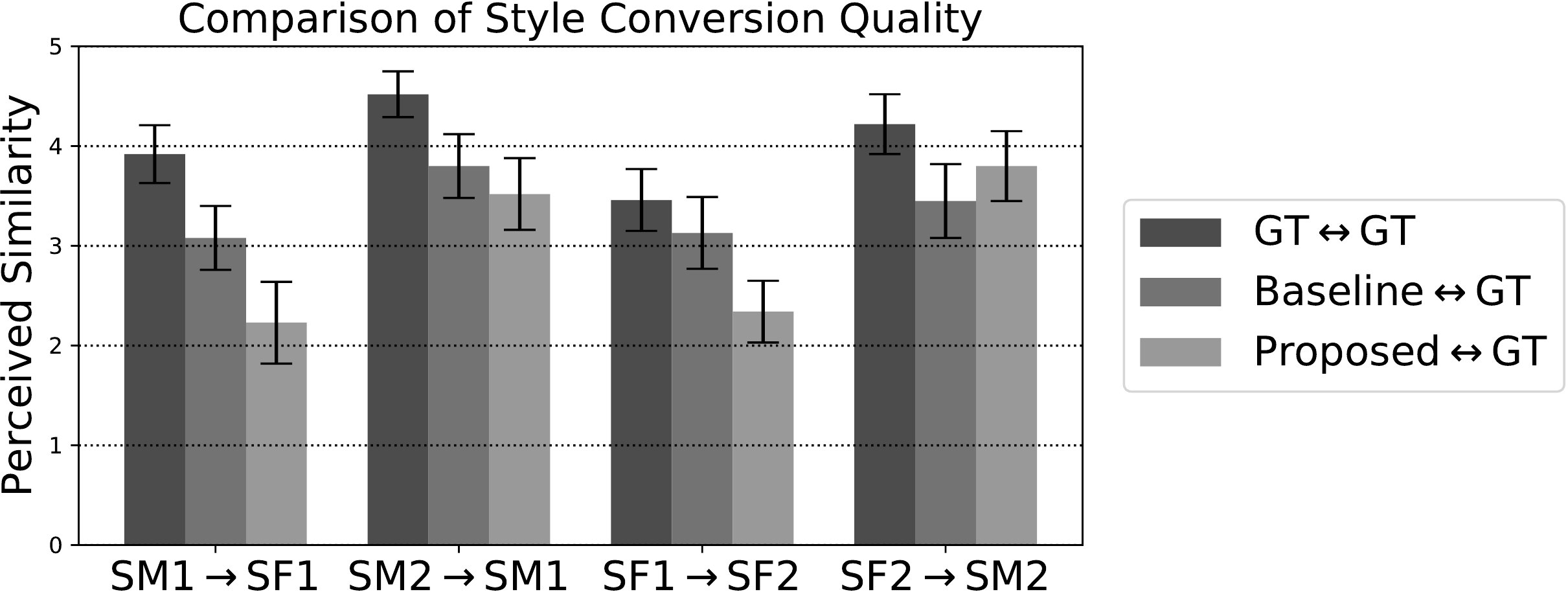}
\end{center}
\caption{Perceived similarity between a converted utterance (baseline or proposed model) and the ground truth (GT) utterance from the intended target of conversion, with 95\% confidence intervals. Higher results are better. GT$\leftrightarrow$GT (perfect conversion) given as a reference. Proposed and baseline conversions are rated similarly.}
\label{fig:results_quality}
\end{figure}

Fig. \ref{fig:results_quality} shows style conversion comparison results. The proposed model performs similar to the baseline in style conversions. The biggest differences between the baseline and the proposed model are in cases where even perfect conversion ($GT \leftrightarrow GT$) is rated relatively lower ($SM1 \leftrightarrow SF1, SF1 \leftrightarrow SF2$). This might indicate inherent style variability for $SF1$ and $SF2$ in the dataset, resulting in lower scores for style conversion.

\subsection{Out-of-Dataset Conversion Quality}
\label{ssec:ood_comp}
Results of Sec. \ref{ssec:baseline_comp} demonstrate that the proposed model performs reasonably well in both naturalness and style conversion for in-dataset speakers. To evaluate out-of-dataset conversion quality, proposed model is trained on 251 speakers in train-clean-100 (Sec. \ref{ssec:dataset}). Speakers in dev-clean (40 in total) are not used for training and are set aside to evaluate the out-of-dataset conversion capability of the trained model. We are unable to compare out-of-dataset conversions to the baseline since the baseline cannot perform such conversions.

Subjective evaluation of conversions between 291 speakers presents a scalability problem. One could pick a few utterances from a source speaker and convert them to an in-dataset and an out-of-dataset target speaker; but subjective test results on such a small set would be highly skewed based on the choice of source and target speakers. For a more robust evaluation, one could pick a set of source speakers, with several utterances from each source, and convert to a set of both in and out-of-dataset targets. This can easily lead to thousands of conversions, each evaluated by several human listeners. Since no prior baseline can perform out-of-dataset conversions, the test would need to be repeated with ground truth utterances for comparison. Time and monetary costs for such testing are prohibitive.

Given these challenges, we opted to use a more scalable method to compare the out-of-dataset target conversion quality to in-dataset target conversions. We trained an i-Vector based speaker identification (SID) model as described  in \cite{kaldiSPID, Povey_ASRU2011}. MFCC  parameters are set as follows: \texttt{$f_s$=16KHz}, \texttt{$frame\:length$=25ms},\texttt{ $frame\:shift$=10ms}, \texttt{$low\:freq$=40}, \texttt{$high\:freq$=7800}, \texttt{$num\:ceps$=20}, \texttt{$vtln\:low$=60}, \texttt{$vtln\:high$=7200}. SID model is trained on all 291 speakers from train-clean-100 and dev-clean datasets (Sec. \ref{ssec:dataset}), with utterances split into training and evaluation sets. SID model reaches 89.5\% top-1 accuracy for utterances in the evaluation set (second to last row in Table \ref{tab:ood_results}). For reference, last row in Table \ref{tab:ood_results} shows the accuracy if the SID model performed random guesses. Conversion model has no knowledge of the SID model and is not specifically trained to perform adversarial attacks on it.

We randomly picked 8 source speakers from train-clean-100 (in-dataset) and 8 from dev-clean (out-of-dataset). 10 utterances per source speaker are converted to the style of 32 target speakers: 16 in-dataset and 16 out-of-dataset (a total of 5120 conversions). Target speakers are split equally between female and male.

Table \ref{tab:ood_results} reports the average top-K accuracy for a converted audio clip to be classified as uttered by the conversion target. Conversion output is rebuilt in raw audio domain using Griffin-Lim \cite{Griffin:1984},  presented to the SID model, and  prediction scores for 291 speakers are sorted. Top-K accuracy measures if the conversion target's score is among the highest ranked K scores.  

First row in Table \ref{tab:ood_results} shows  average classification accuracy of conversions for 16 in-dataset target speakers. When presented to the SID system, 9.7\% of the converted utterances are predicted to come from the intended target (among 291 potential target speakers). Average accuracy for out-of-dataset targets is lower at 4.8\%, but significantly higher than random chance (0.3\%); demonstrating that $\pmb{FE}$ can generate reasonable embeddings (\textit{style}) for new speakers.

Table \ref{tab:granular} reports more granular results based on in-dataset/out-of-dataset status of speakers, as well as target speakers' gender. Out-of-dataset source speakers have only slightly lower accuracy. This is likely because downsampling path of $\pmb{G}$ learns to discard the \textit{style} of the source speaker while keeping the \textit{content}; hence out-of-dataset source speakers' styles do not significantly impact the conversion.

\begin{table}[t!]
\centering
\caption{Quantitative Comparison of Style Conversion Quality for In-Dataset and Out-of-Dataset Target Speakers}
\label{tab:ood_results}
\resizebox{0.9\columnwidth}{!}{
\begin{tabular}{c|c|c|c|c|c}
\multicolumn{1}{ c }{ Target} & \multicolumn{5}{| c }{Accuracy (Percent)}\\
\multicolumn{1}{ c }{Speakers} & \multicolumn{1}{| c |}{Top-1} & Top-3 & Top-5 & Top-10 & Top-20 \\
\toprule[.2em]
 In-Dataset & 9.7 & 19.8 & 27.5 & 39.6 & 54.0\\
 Out-of-Dataset & 4.8 & 11.5 & 15.5 & 23.9 & 35.5\\
 \toprule[.15em]
 SID Eval. Set & 89.5 & 95.3 & 96.5 & 98.0 & 98.9\\
 Chance & 0.3 & 1.0 & 1.7 & 3.4 & 6.8
\end{tabular}
}
\end{table}

\begin{table}[t!]
\centering
\caption{Granular Comparison of Style Conversion Quality with Different Speaker Genders and/or Datasets}
\label{tab:granular}
\resizebox{1\columnwidth}{!}{
\begin{tabular}{c|c|c|c|c|c|c}
\multicolumn{1}{ c |}{ Source} & \multicolumn{1}{ c }{ Target} & \multicolumn{5}{| c }{Accuracy (Percent)}\\
\multicolumn{1}{ c |}{Speakers} & \multicolumn{1}{ c }{Speakers} & \multicolumn{1}{| c |}{Top-1} & Top-3 & Top-5 & Top-10 & Top-20 \\
\toprule[.2em]
  \multirow{4}{1cm}{\centering In Dataset}&Female, In & 10.0 & 20.6 & 28.1 & 40.8 & 57.5\\
                           &Male, In & 12.3 & 21.9 & 30.8 & 45.5 & 57.8\\ 
                           &Female, Out & 4.4 & 11.4 & 14.7 & 24.1 & 37.5\\
                           &Male, Out & 6.7 & 14.7 & 18.0 & 25.6 & 35.2\\
  \toprule[.1em]
  \multirow{4}{1cm}{\centering Out of Dataset}&Female, In & 7.2 & 17.2 & 25.2 & 36.6 & 52.2\\
                           &Male, In & 9.2 & 19.7 & 25.8 & 35.6 & 48.6\\ 
                           &Female, Out & 3.0 & 8.9 & 12.7 & 21.6 & 35.0\\
                           &Male, Out & 5.2 & 11.1 & 16.7 & 24.2 & 34.5\\
\end{tabular}
}
\end{table}

\section{DISCUSSION}
\label{sec:discussion}
In this paper, we describe a non-parallel, almost unsupervised, Cycle-GAN based voice conversion method that can perform conversions between speakers that the model was never trained on. This is enabled by a unique feature extractor block that produces speaker embeddings for new speakers. Subjective tests show that style conversion quality is comparable to the state of the art, which can only perform in-dataset conversions. Out-of-dataset conversion quality of the proposed model is compared to the in-dataset quality using a quantitative method based on an independently trained speaker identification model. Future work includes improving the naturalness of the converted speech by employing a vocoder, and improving out-of-dataset conversion quality by training on a larger set of speakers to enhance the generalization capability of the feature extractor.

\bibliographystyle{IEEEtran}

\bibliography{mybib}

\end{document}